\documentclass[acmsmall,10pt]{acmart}
\usepackage{supertech-acm}
\usepackage{flushend}
\usepackage{xspace,balance}

\usepackage[linesnumbered,ruled,figure]{algorithm2e}

\notesfalse  

\newcommand{\work}{T_1} 
\newcommand{\spa}{T_\infty}

\newcommand{\suspendWidth}{U}

\newcommand{\spawn}{\texttt{spawn}\xspace}
\newcommand{\sync}{\texttt{sync}\xspace}
\newcommand{\cilkfor}{\texttt{cilk\_for}\xspace}
\newcommand{\createF}{\texttt{fut-create}\xspace}
\newcommand{\getF}{\texttt{get}\xspace}
\newcommand{\putF}{\texttt{put}\xspace}
\newcommand{\getFNoSpace}{\texttt{get}}
\newcommand{\putFNoSpace}{\texttt{put}}

\newcommand{\ALG}{ProWS\xspace}

\newcommand{\SYS}{Cilk-L\xspace}

\newcommand{\IOT}{I/O thread\xspace} 
\newcommand{\IOTS}{I/O threads\xspace}
\newcommand{\ioF}{\texttt{io\_future}\xspace}
\newcommand{\cilkWrite}{\texttt{cilk\_write}\xspace}
\newcommand{\cilkRead}{\texttt{cilk\_read}\xspace}

\newcommand{\cilkWriteNoSpace}{\texttt{cilk\_write}}
\newcommand{\cilkReadNoSpace}{\texttt{cilk\_read}}
\newcommand{\sysWrite}{\texttt{write}\xspace}
\newcommand{\sysRead}{\texttt{read}\xspace}
\newcommand{\commQ}{communication queue\xspace}
\newcommand{\epollNoSpace}{\texttt{epoll}}
\newcommand{\epoll}{\epollNoSpace\xspace}
\newcommand{\eventfdNoSpace}{\texttt{eventfd}}
\newcommand{\eventfd}{\eventfdNoSpace\xspace}

\newcommand{\benchName}{\texttt{map-reduce}\xspace}

\title{Reduced I/O Latency with Futures}

\author{Kyle Singer}
\affiliation{\institution{Washington University in St. Louis}}
\email{kdsinger@wustl.edu}

\author{Kunal Agrawal}
\affiliation{\institution{Washington University in St. Louis}}
\email{kunal@wustl.edu}

\author{I-Ting Angelina Lee}
\affiliation{\institution{Washington University in St. Louis}}
\email{angelee@wustl.edu}

\makeatletter
\let\@authorsaddresses\@empty
\makeatother

\setcopyright{none}
\renewcommand\footnotetextcopyrightpermission[1]{} 
\setcitestyle{authoryear}

\begin{document}

 \begin{abstract}

   Task parallelism research has traditionally focused on optimizing
   computation-intensive applications.  Due to the proliferation of
   commodity parallel processors, there has been recent interest in
   supporting interactive applications.  Such interactive applications
   frequently rely on I/O operations that may incur significant
   latency.  In order to increase performance, when a particular
   thread of control is blocked on an I/O operation, ideally we would
   like to hide this latency by using the processing resources to do
   other ready work instead of blocking or spin waiting on this I/O.
   There has been limited prior work on hiding this latency and only
   one result that provides a theoretical bound for interactive
   applications that use I/Os.  

   In this work, we propose a method for hiding the latency of I/O
   operations by using the futures abstraction. We provide a
   theoretical analysis of our algorithm that shows our algorithm
   provides better execution time guarantees than prior work.  We also
   implemented the algorithm in a practically efficient prototype
   library that runs on top of the Cilk-F runtime, a runtime system
   that supports futures within the context of the Cilk Plus language,
   and performed experiments that demonstrate the efficiency of our
   implementation.

\end{abstract}

\maketitle

\thispagestyle{empty}

\secput{intro}{Introduction}

With the prevalence of multicore processors, task parallelism has become
increasingly popular.  With task parallelism, the programmer expresses the
\emph{logical} parallelism of the computation and let an underlying runtime
system handles the necessary load balancing and synchronizations.  Modern
parallel platforms that implement task parallelism include but are not limited to
OpenMP~\cite{OpenMP13}, Intel TBB~\cite{IntelTBBManual}, various dialects of
Cilk~\cite{IntelCilkPlusLangSpec13, LeeBoHu10, Leiserson10, DanaherLeLe06} and
Habanero~\cite{BarikBuCa09, CaveZhSh11}, X10~\cite{CharlesGrSa05}, and Java
Fork/Join framework~\cite{Lea00}.  These platforms often schedule parallel
computations using work stealing, which provides provable bounds on execution
time~\cite{BlumofeLe94, BlumofeLe99, AroraBlPl98, AroraBlPl01}, good space
bounds~\cite{BlumofeLe99}, good cache locality~\cite{AcarBlBl00, AcarBlBl02},
and allows for an efficient implementation~\cite{FrigoLeRa98}.

Research on task parallel platforms has traditionally focused on
optimizing for compute-intensive and throughput-oriented applications,
such as ones found in the domain of high-performance and scientific
computing.  Multicore processors have become commonplace and used in
personal computers and servers, however, and a fundamental component
of desktop software is its frequent interactions with the external
world, done in the form of input/output (I/O), such as obtaining user
input through key strokes or mouse clicks, waiting for a data packet
to arrive on a network connection, or writing output to a display
terminal or network.

The classic work stealing algorithm does not account for I/Os.  I/O
operations are typically done via low-level system libraries (e.g.,
the GNU C library) or through system calls provided by the Operating
Systems (OS).  While one can directly invoke functions provided by
these libraries within a task parallel program, doing so has
performance implications.  In particular, when a \defn{worker thread}
--- surrogate of a processing core managed by the scheduler ---
encounters an I/O operation, it can block for an extended period of
time, leaving one of the physical cores underutilized.  \footnote{Low-level
system support for \defn{asynchronous} (non-blocking) I/O exists, but
resuming the context on getting the I/O completion (typically a
signal) can be complex.}

In this work, we design a scheduler that \emph{hides} the I/O latency
--- when a worker encounters a blocking I/O, it simply suspends the
current execution context and works elsewhere in the computation.
When the I/O completes, some worker (not necessarily the worker that
suspended it) picks up the suspended context and resumes it.
Moreover, the programming model seamlessly integrates both blocking
and nonblocking I/Os into the task parallel programming model.
Finally, the scheduler provides provably good performance bounds
and efficient implementation.

As far as we know, only one prior result provides provably efficient
scheduling bound of task parallel programs with I/Os.~\citet{MullerAc16} present a cost model for reasoning about
latency incurring operations (such as I/Os) in task parallel programs.
In their work, given a computation with \defn{work} $\work$ --- the
total computation time on one core --- and \defn{span}\footnote{The
  term span is sometimes called ``critical-path length'' and
  ``computation depth'' in the literature.} $\spa$--- the execution
time of the computation on infinitely many cores --- the scheduler
executes the computation in expected time
$O(\work / P + \spa \suspendWidth (1 + \lg \suspendWidth))$, where the
$\suspendWidth$ is the maximum number of latency incurring operations
that are logically in parallel.  Their bound is \defn{latency-hiding}
in that, the latencies of I/O only appear in the span term and not the
work term.  If no latency-incurring operations are used, their bound
is asymptotically equal to the standard work stealing bound of
$O(\work / P + \spa)$.

In this work, we improve the latency-hiding bound by using a
scheduling algorithm based on \ALG~\cite{SingerXuLe19}, a recently
developed work-stealing scheduler that efficiently supports futures.
We implement I/O operations seamlessly within task parallel code using
futures while getting nearly asymptotically optimal completion time.
In particular, we were able to prove that our latency-hiding scheduler
provides an execution time bound of $O(\work / P + \spa \lg P)$ in expectation; this
bound is independent of the number of I/Os in the
system.  In particular, compared to the standard work-stealing bound,
it just has an additional term of $\lg P$ on the span term.  This
implies that while the standard work-stealing scheduler provides
linear speedup when $\work/\spa = \Omega(P)$, our scheduler provides
linear speedup when $\work/\spa =  \Omega(P \lg P)$.  
\ALG has the same bound, but the analysis does not directly apply here
due to the latency of the I/Os.  We extend \ALG's bound to futures
with I/Os.

Once we extend the analysis, we essentially inherit the
bound on ``deviations'' from \ALG.  Intuitively, deviations are points
at which the parallel execution of a program differs from its
sequential execution.  ~\citet{SpoonhowerBlGi09} argue
that the number of deviations provides a good metric for evaluating
practical performance because it is highly correlated to scheduling
overheads and cache misses during parallel executions.  \ALG (and we)
guarantee that the expected number of deviations is
$O((P \lg P + m_k) \spa)$ where $m_k$ is the total number of futures
which are logically in parallel.  

The high-level intuition on why using futures to do I/Os and then
using \ALG to schedule these futures provides better bounds is as
follows: The work-stealing algorithm by Muller and Acar is
\defn{parsimonious} --- a worker never steals unless it runs out
of work to do.  In contrast, \ALG's and our work-stealing algorithm is
\defn{proactive} --- whenever a worker encounters a blocking I/O
operation, it suspends the entire execution context and finds
something else to do by work stealing.  This behavior may seen
counter-intuitive since it potentially increases the number of steal
attempts.  It turns out, however, that that by carefully managing
deques, one can amortize the steal cost against the work term
sometimes, thereby obtaining a better bound.  More importantly, we can
get good bounds on deviations for the following reason: In the
earliest result on deviations,~\citet{AcarBB02} related the
number of deviations to the number of steal attempts for fork-join
programs.  However, this relationship does not hold in parsimonious
work-stealing if the program uses unstructured blocking operations
like futures or I/Os making it difficult to bound the number of
deviations.  In proactive work-stealing, we can again bound deviations
using the number of steals, allowing us to bounds deviations.

Our prototype system \SYS is based on Cilk-F~\cite{SingerXuLe19}, an
extension of Intel Cilk Plus~\cite{IntelCilkPlus13} that supports
futures and implements \ALG.  \SYS is able to defines a special type
of futures, called \defn{IO futures}, which utilize the parallelism
abstraction provided futures to schedule I/Os in a latency-hiding manner
which is composable with the rest of parallel constructs supported in
Cilk-F (\spawn, \sync, \createF, and \getF; we will briefly discuss these
in \secref{prelim}).  When a worker invokes an I/O operation using IO
futures, a handle is returned, and the I/O can be done either
\defn{synchronously} by calling \getF on the handle immediately, or
\defn{asynchronously}, calling \getF at a later time when the result
is needed in order for the control to proceed.

We empirically evaluated \SYS with microbenchmarks that interleave
compute-intensive kernels with operations that incur I/O latencies.
The empirical results indicate that, \SYS is effective at latency
hiding.  When we compare the execution times of \SYS with the
``idealized' execution times (where the I/O does not incur latency),
we find that \SYS incurs little overhead, indicating the the I/O
latencies are mostly hidden and occur in the background.  In order to
support future I/O, \SYS necessarily needs to incorporate additional
system support for scheduling I/O asynchronously.  We also provide an
detailed breakdown of overhead.

\paragraph{Summary of contributions:}
\begin{closeitemize}

\item We extend the scheduling algorithm in Cilk-F to incorporate the
  latency-hiding cost model, and show that with I/O latency, the
  algorithm can schedule the computation in time
  $O(\work / P + \spa \lg P)$ on $P$ cores, independent of the number
  of I/O operations active in parallel.  This bound is an improvement
  over the prior state-of-the-art by Muller and Acar.  Since
  $\max\{\work/P, \spa\}$ is a lower bound on the execution of this
  program on $P$ processors, this bound is nearly asymptotically
  optimal except for the $\log P$ overhead on the span.  Moreover, our
  algorithm provides bounds on stack space and deviations, whereas the
  algorithm by Muller and Acar does not (\secref{analysis}).

\item We developed \SYS by extending Cilk-F to incorporate support for
scheduling I/O in a latency-hiding way.  By utilizing the abstraction of
futures, one can perform asynchronous I/Os in task parallel code in a way that
is composable with other parallel constructs (\secref{system}).

\item We empirically evaluated \SYS using microbenchmarks.  The empirical
results indicate that \SYS hides I/O latencies effectively and incurs little
scheduling overhead in doing so (\secref{eval}).

\end{closeitemize}

\punt{
The rest of the paper is organized as follows. \secref{prelim} provides the
necessary background. \secref{system} discusses \SYS and its extension on top
of Cilk-F. \secref{analysis} briefly explains the algorithm used in Cilk-F and
proves its performance bounds.  \secref{eval} presents the empirical
evaluation.  \secref{related} discusses related work and \secref{conclusion}  
offers concluding remarks.
}

\secput{prelim}{Preliminaries}

This section provides the necessary background.  We first discuss the
syntax and semantics for the parallel control constructs supported by
Cilk-F~\cite{SingerXuLe19} and how one can represent a computation
expressed with these keywords abstractly as a DAG.  We then discuss
how a \emph{parsimonious} work stealing runtime schedules the
computation assuming no latency-incurring operations are present.

\paragraph{Parallel Control Constructs:}
Cilk-F, and by extension \SYS, support a small set of parallel control
constructs: \spawn, \sync, \createF, and \getF.\footnote{The keyword
\cilkfor also exists to indicate parallel loops, but it is just a
syntactic sugar that translates to binary spawning of iteration space
using \spawn and \sync.}  In Cilk-F, these keywords operate at the
level of function calls.  When a function $F$ \defn{spawns} off a
function $G$ by prefixing the call with the \spawn keyword, $G$ may
execute in parallel with the \emph{continuation} of $F$ (the
statements after the \spawn).  The keyword \sync is the counterpart of
\spawn; it indicates that control cannot pass beyond the \sync
statement until all previously spawned children have returned.  In
Cilk-F, there
is an implicit \sync at the end of every function, ensuring that all
children spawned via \spawn return before this function returns.

The keyword \createF works in a similar fashion as \spawn.  When a
function $F$ spawns off a function $G$ by prefixing the call with the
\createF call, $G$ may execute in parallel with the
\emph{continuation} of $F$.  Unlike \spawn, however, the execution of
a \sync has no effect on \createF.  The control \emph{can} pass beyond
\sync even if a function previously spawned off via \createF has not
returned.  Moreover, a \createF returns a \defn{handle} $h$, which is
an object that the execution of $G$ is associated with.  The handle
can later be used to ensure termination of $G$ and retrieve its
result.  In particular, when $G$ finishes execution, the last
instruction is implicitly a \putF call which puts the result of $G$
into $h$ and marks the future as ready.  By invoking \getF on the
handle, the control cannot pass beyond the \getF until the execution
of $G$ terminates and the future is marked as ready.

\paragraph{Execution DAG:}
Parallel computations generated by programs written with these
primitives can be represented using a directed acyclic graph
(DAG). Vertices of the DAG represent a unit time computation
task\footnote{This is an assumption of convenience --- longer
operations can be represented as a chain of unit time operations.} and
edges represent dependences between nodes.  We make the
standard assumptions: there is a single root node and the
out-degree is at most 2.

We classify nodes into a few different categories.  Regular nodes are
simply computation nodes.  A \defn{spawn} node executes a \spawn and
has two children --- the left child is the first node of the spawned
function and the right child is the continuation node.  A \defn{join}
node represents the continuation after a \sync call and has multiple
parents --- \sync node itself and the last nodes of all the functions
being synced.  The \createF keyword behaves similarly to \spawn and
generates a \defn{future spawn} node that has two children: the left
child is the first node of the future task and right is the
continuation.  A \defn{future join} node is the node immediately after
the invocation of \getF and has two parents --- the \getF node (called
the \defn{local} parent) and the \defn{future put node} --- the last
node of the corresponding future task that puts the result of the
future in the future handle.

We say that a node is \defn{ready} or \defn{enabled} if all its
predecessors have executed.  The \defn{work} of the computation DAG is
the total number of nodes in the DAG and is represented by $T_1$ ---
it is the total time to execute the DAG on 1 processor.  The span of
the weighted DAG is the longest path in the DAG and is represented by
$T_\infty$.

\paragraph{Parsimonious Work Stealing:}
As mentioned in \secref{intro}, parsimonious work-stealing works by
doing local work first.  In computations with no latency forming
blocking operations, each worker maintains a single \defn{double
ended-queue} (or \defn{deque}) of ready nodes.  For the most part, a
worker operates on its deque. In particular, when a worker finishes
executing a node, it may enable 0, 1 or 2 of its children.  If it
enables one child, the worker next executes the child.  If it enables
two children, it puts the right child on the deque and executes the
left child.  If it enables no children, it pops the bottom node from
its deque and executes that node.  Only when a worker runs out of work
(its deque comes empty), does it turn into a \defn{thief}.  At this
time, it randomly chooses a \defn{victim} to steal work from.  Upon
steal, the thief steals the ready node from the top of the victim's deque
and executes it.  If victim deque has no ready nodes, then the worker
tries another random steal.  




\secput{system}{The System Implementation}

This section describes \SYS, a prototype system that extends
Cilk-F~\cite{SingerXuLe19} to incorporate support for performing I/Os
with latency hiding.  The I/O support in \SYS consists of two main
components: the \defn{IO futures} library and runtime
support for doing asynchronous I/Os.  We first discuss the programmer
API for using IO futures, its implementation, and then the runtime
support for asynchronous I/Os.  Since the I/O operations are typically
supported via low-level system libraries and by the underlying
Operation System, currently \SYS only targets Linux platform and
utilizes various file-I/O related facilities from Linux.

\subsection*{The IO Futures Library}

%
We use an example to illustrate the programming API provided by the library.
\figref{example} shows the distributed map-and-reduce example used by~\citet{MullerAc16}.  Our microbenchmark effectively uses the same
parallel structure to generate workload with I/O latencies; doing so allows us
to indirectly compare our results with the empirical results
in~\citet{MullerAc16} (discussed in \secref{eval}).

In this example, the function \texttt{distMapReduce} takes in five parameters:
$f$, $g$, $id$, $lo$, and $hi$.  The computation works as follow.  The code
obtains a set of input values from $n = hi - lo$ different network
connections.  In Linux, all I/O devices are presented as files, including
network connections, which allows for a uniform interface for performing
I/O~\cite[Chp. 10]{BryantOH15}.  The call to \texttt{openConnection} in
\liref{connect} abstracts away the sequence of steps to open a network
connection, which returns a file descriptor representing the connection once
it's open.  For each value $x$ in the set the code applies the map function
$f(x)$, and then combine the resulting values of applying $f(x)$ using a
binary reduction operation $g$.  

The IO futures library exposes one data type to the programmer, the
handle for IO futures \ioF, and two I/O functions, \cilkRead and
\cilkWrite.  The \cilkRead and \cilkWrite functions are analogous to
the Unix \sysRead and \sysWrite system calls, except that they are
\defn{asynchronous}, i.e., non-blocking.  Upon calling, both functions
return an \ioF handle representing the on-going I/O operation once
it's set up but the function itself does not block on the I/O.
However, when the result is needed in order to proceed, the programmer
can invoke \getF on the \ioF handle.  Like \getF on an ordinary
future, control cannot pass beyond \getF on the returned \ioF until
the corresponding I/O operation completes.\footnote{The worker itself
  does not block when this happens --- the worker takes actions
  according to the proactive work-stealing strategy described in the
  next section.}
\begin{algorithm}
\footnotesize

\SetKwProg{Fn}{Function}{}{end}
\SetKwFunction{GetFuture}{\getFNoSpace}
\SetKwFunction{PutFuture}{\putFNoSpace}
\SetKwFunction{CilkWrite}{\cilkWriteNoSpace}
\SetKwFunction{CilkRead}{\cilkReadNoSpace}
\SetKwFunction{OpenConnection}{openConnection}
\SetKwFunction{DistMapReduce}{distMapReduce}

\SetKw{Break}{break}
\SetKw{Continue}{continue}
\SetKw{Spawn}{\spawn}
\SetKw{Sync}{\sync}

\SetInd{0.5em}{0.5em}
\SetNlSkip{0.5em}
\SetKwComment{tcp}{/\hspace{-0.2em}/}{\hspace{0.5em}}

\Fn{\DistMapReduce{$f$, $g$, $id$, $lo$, $hi$}} {
   $n \gets hi - lo$\;
   
   \lIf(\tcp*[h]{return identity.}){$n = 0$}{\Return $id$}
   \ElseIf{$n = 1$}{
        char $buf$[NBYTES] \tcp*[l]{buffer for input data.}
   	
        $fd \gets$ \OpenConnection{$lo$}
            \tcp*[h]{open network connection.}
            \label{li:connect}
    
   	\ioF $fut \gets$ \cilkReadNoSpace($fd$, $buf$, NBYTES)
                            \label{li:read}
   	
   	\GetFuture{$fut$};
   	
   	\Return $f$($buf$);
   } \Else {
	$mid \gets (lo+hi)/2$;
	
	$r1 \gets$ = \Spawn \DistMapReduce{$f$, $g$, $id$, $lo$, $mid$};
	
	$r2 \gets$ \DistMapReduce{$f$, $g$, $id$, $mid$, $hi$};
	
        \Sync;
	
	\Return $g$($r1$, $r2$)\label{li:g}
}
}
\caption{Distributed map and reduce pseudo code.}
\label{fig:example}
\end{algorithm}

%
Every call to a \SYS I/O function first creates an \ioF to represent
the non-blocking I/O request, and bundles it with the corresponding
data required to carry out the I/O request (such as the file
descriptor $fd$ and the buffer $buf$ to store input).  This data
bundle is then inserted into a lock-free
single-producer/single-consumer queue, which we refer to as the
\defn{\commQ}, to be processed by the runtime.  The \ioF is then
returned to the caller.  If the user needs the result from the future
or wants to ensure that it has completed, it can perform a \getF on
this handle \ioF.  The instruction immediately after this \getF (the
continuation of \getF, in other words, the future join node) can not
execute until the I/O has completed.

\subsection*{Runtime Support for Hiding I/O Latencies}

At runtime startup, normally Cilk-F creates $P$ persistent threads for $P$
workers.  In \SYS, $2P$ persistent threads are created --- for every worker a
corresponding \defn{\IOT} is created, and this persistent thread is pinned to
the same core as its worker.\footnote{If the hardware has hyperthreading
enabled, \SYS pins them to separate hardware threads (hyperthreads) associated
with the same physical core.}  The \IOT is only used to process I/O requests
(via the IO futures library) generated by the worker's execution of user code.
Thus, in the library implementation described above, the \commQ is used as
means for the worker thread to communicate I/O requests to its \IOT.

When an \IOT runs, it dequeues items from the \commQ and attempts to perform
an I/O operation as soon as it is received.  If an I/O operation cannot be
completed immediately (e.g., the next package has not arrived on the network
channel yet), however, we would like to put the request aside and process it
later when the I/O device becomes ready (has more input to be consumed).  

In order to describe how the actual mechanism works, we need to
briefly discuss how I/O works on Linux.  As mentioned earlier, any I/O
device on Linux is represented as a file descriptor.  Obtaining an
input (read) from a file descriptor is effectively copying data from
the corresponding device into memory (e.g., the $buf$ in the example).
If the device is not ready to be read (e.g., the next package has not
arrived on the network channel yet), the system call \sysRead will
block.  One could make the system call with the correct flag so that
the system call would simply return instead of blocking, with a return
value indicating input not ready.  However, in this case, we must make the
system call to check back periodically in order to know when a device
becomes ready.

On possibility is to periodically wake up the \IOT and have it poll the device
via non-blocking \sysRead.  This scheme is not ideal, as a system call can be
expensive.  Moreover, if the device is not ready, checking would simply cause
the \IOT to take up processor cycles that could be better used by its worker
working on the actual computation.  Thus, we would like to avoid the periodic
wake up and the unnecessary system calls.  Ideally, we would like the \IOT to
simply sleep and not use any processor cycles \emph{unless} one of the
following conditions happen: a) one of the file descriptors with pending
operations becomes ready; or b) its worker inserts a new I/O request into the
\commQ.

To achieve part a), we use the Linux \epoll~\cite{LinuxManPageEpoll19}
facility which allows the \IOT to monitor a set of file descriptors (an \epoll
set).  Adding a file to be monitored is an $O(\lg n)$ operation, where $n$ is
the number of file descriptors currently in the \epoll set.  The \IOT can go
to sleep by calling \texttt{epoll\_wait}, and it will be woken up when one of
the monitored file descriptors become ready.  Determining which file
descriptors monitored has become ready is a $O(1)$ operation --- adding a file
descriptor to the \epoll set registers a callback with the file's underlying
system driver; this callback will move the file into a ready list and wake the
monitoring thread when I/O on that file becomes possible.  Once the \IOT is
woken up, it can query \epoll to obtain the list of ready file descriptors,
which allows the \IOT to determine which pending I/O operations can continue.
In summary, each \IOT maintains its own \epoll set.  When an \IOT receives an
I/O request but the corresponding file descriptor is not ready, the \IOT adds
the file descriptor to its \epoll set to be monitored.  Once an \IOT has
processed all I/O requests in the \commQ, it goes to sleep via
\texttt{epoll\_wait}.  Doing so achieves part a).  

One last piece puzzle is how one to avoid having the \IOT check the \commQ
periodically and yet still allow submitted I/O requests to be processed
quickly whenever it is received.  We solve this by using an event wait/notify
mechanism called \eventfd provided by Linux~\cite{LinuxManPageEventfd19}.  The
\eventfd mechanism is used to create a file descriptor that can read by a \IOT
and written to by its worker.  This file descriptor can be opened with
semaphore-like semantics, in which writes will increment a backing counter and
reads will decrement the same counter.  When used with \epoll, a write to an
\eventfd file descriptor will cause the \IOT to wake up whenever the backing
counter is incremented from 0 to 1. By writing to an \eventfd file descriptor
associated with a \commQ whenever an I/O operation is enqueued, and by
symmetrically reading from the same file descriptor whenever an operation is
dequeued, \epoll can also be used to monitor the state of the \commQ.  Thus,
we use this  combination of \eventfd and \epoll to achieve part b).

By combinations of these mechanism, we achieve the effect that, an \IOT is
only woken up and take up processor cycles when either there is a new I/O
requests from the worker or when one of the previously processed I/O that was
pending now becomes ready.  When a \IOT completes an I/O operation, it
performs a \putF on the corresponding \ioF handle.  From its worker's
perspective, a call to \getF can cause the current execution to suspend, but
the worker would just go find something else to do.  \SYS schedules the
execution of the IO futures just as how Cilk-F schedules ordinary futures,
which we briefly review in \secref{analysis}.

\secput{analysis}{Algorithm and Analysis}

In this section, we will describe how to represent a program with I/Os
abstractly, the high level scheduling algorithm, and the runtime analysis.
For scheduling, we will essentially use \ALG, the proactive work-stealing
algorithm described by~\citet{SingerXuLe19}.  The algorithm
described in that paper schedules programs with futures in a time and space
efficient manner.  For completeness, we will briefly describe the algorithm
here.  However, the analysis in that paper handles futures but not I/Os.  Here
we will show how that analysis can be extended to also appropriately handle
I/O latencies.

\subheading{Execution DAG}

We will extend the model from \secref{prelim} and add weighted edges
in a manner similar to~\citet{MullerAc16}.  
In our model, I/O operations are performed within future tasks.  The
invocation of an I/O function (\cilkRead and \cilkWrite) creates an \ioF
implicitly, sets up the necessary data for the I/O request, inserts the
request into the \commQ (discussed in \secref{system}), and returns.  We will
call the last node of this future task before it returns the \defn{I/O setup
node}.  However, unlike in non-I/O future tasks, this future itself is not
ready.  The future is ready when the \IOT and executes \putF upon the I/O
completion --- we will call the put node of an I/O future an \defn{I/O put}
node.  We will have a \defn{heavy} edge between the I/O setup node and the
corresponding I/O put node --- the weight on this edge represents the amount
of time it took for the I/O to complete.  All other edges are \defn{light} in
that they have weight of one.

We can define work and span --- the work is unchanged: it is the total
number of nodes in the DAG.  Therefore, it is unaffected by the
latencies on the edges.  The span of the weighted DAG is the longest
weighted path in the DAG and is the only parameter affected by the
latencies.

Again a node is \defn{ready} if all its predecessors have executed,
except for the I/O put node.  An I/O put node is \defn{suspended} once
its predecessor (the corresponding I/O setup node) finishes executing.
If $\ell$ is the weight of the incoming edge to the put node, it
remains suspended for $\ell$ time steps.  After these $\ell$ time
steps, it is considered to have finished executing since the I/O
thread will write the result into the future handle after these $\ell$
time steps.  This definition of suspension of a put node is simply for
the ease of analysis and has no impact on the scheduler since the put node is
executed by an \IOT and not by the worker thread.  

\subheading{Proactive Work Stealing}

We use~\citet{SingerXuLe19}'s proactive work stealing
scheduler unchanged.  The main difference between proactive and
parsimonious work stealing is the handling of a blocked future get.
In parsimonious work stealing, when a worker's current node executes a
\getF and the future is not ready, the subsequent future join node is
not enabled.  Therefore, the current node enables 0 children and (as
described in \secref{prelim}) the worker pops the next node from the
deque and continues working on it. ~\citet{MullerAc16}'s algorithm is
a variant of this --- when a worker blocks on an I/O, it pops the next
node off its deque and keeps working on it.

A proactive work-stealer behaves differently on executing a \getF
where the future handle $h$ which is not ready.\footnote{There are
other circumstances where the execution of the node enables no other
nodes such as when a worker returns from a spawned or future function --- in
all these circumstances proactive work stealng behaves as the parsimonious one
and pops the bottom node from its deque.} instead of popping the next node
from its active deque $d$, the worker work steals.  In particular, the worker
(1) marks the current deque \defn{suspended}; (2) it randomly picks another
worker and donates this suspended deque with this worker; and (3) allocates a
new active deque $d'$ for itself and randomly work steals.  When the handle
$h$ become ready (the future finishes), then the corresponding put node marks
the deque $d$ \defn{resumable} and pushes the future join node to the bottom
of $d$.

Therefore, in a proactive work-stealing scheduler, each worker $p$ has
potentially many deques.  One of these is \defn{active} --- this is the deque
the worker is currently working on.  In addition, it many have many suspended
and resumable deques --- collectively, the suspended and resumable deques are
called the worker $p$'s \defn{inactive} deques.  In addition, any suspended
deques that have no ready nodes are \defn{unstealable}; all other deques are
\defn{stealable}.  The reason for this distinction is that unstealable deques
have no ready nodes, so stealing from them is a waste of time.  Note that any
resumable deques with no ready nodes are simply de-allocated.  However, a
suspended deque $d$ with no ready nodes cannot be deallocated for the
following reason.  Deque $d$ is suspended since some \getF executed, but the
corresponding future has not completed.  When this future completes, the
corresponding put node will enable the future join node and push it at the
bottom of $d$ and mark it resumable.  Therefore, if we deallocate it, we would
not have a targeted place to push this future join node.  

A steal attempt also works slightly differently compared to traditional work
stealing.  When work stealing, a thief first picks a random victim and then
picks a random stealable deque to steal from among the deques that the victim
has.  If the target deque is suspended, then the worker simply steals the top
node from the deque.  If the deque has no more ready nodes, then this deque is
marked unstealable.  There are additional details on how to handle resumable
deques in order to get the correct bounds on running time and deviations ---
however, these details do not change in our analysis and we refer the reader
to~\citet{SingerXuLe19} for those details.  

The important bits from the perspective of our understanding are the
following: (1) Every worker has potentially many deques: one deque is
active, and there are potentially many inactive deques (either
suspended or resumable and some of the suspended deques may be
unstealable); and (2) due to random throws when the deques are
suspended, all workers have approximately equal number of deques.  We
will use these two facts in the analysis.




\subheading{Analysis}

The analysis of the proactive work-stealing scheduler is, to a large extent,
an extension of the analysis by~\citet{SingerXuLe19} (henceforth, we
will call them \defn{SXL}) with proper accounting for latency edges.  Muller
and Acar do account for latency on edges, but do not use futures for I/O's,
use parsimonious work stealing, and do not rebalance deques between workers.
Therefore, the running time on $P$ processors is $O(\work/P + \spa U (1+\lg
U))$, where $U$ is the \defn{maximum suspension width} --- the number of I/O's
that can be pending at the same time in the DAG.  There is no bound on the
number of deviations.  The way they handle the potential function in order to
hide the latency is a little different from our method.

Here, we are using parsimonious work stealing and want to get a running time
bound of $O(\work/P + \spa \lg P))$ and the deviation bound of $O((P\lg P +
m_k)\spa)$ where $m_k$ is the total number of futures logically in parallel.
For the special case where all futures are I/O futures, $m_k=U$.\footnote{SXL
provide separate deviation bounds for structured and general futures and they
both carry over.  If all our I/Os are done using structured futures, then the
bound is better than this --- here we only provide the general bound.} Their
analysis doesn't work out of the box, however, since it does not consider the
latency on heavy edges.  Therefore, here, we will rely on the lemmas proved in
that paper, but modify the potential function in order to handle the heavy
edges appropriately.

In general, in work stealing, a worker is always either working or
stealing.  The main point of the analysis is to bound the total number
of steal attempts, say by $X$.  Since the total work is $\work$, the
total running time is $(\work + X)/P$.  In addition, bounding the
total number of successful steals also gives us a bound on the total
number of deviations (for proactive work stealing, though not for
parsimonious work stealing).

\paragraph{\ALG Potential Function and Analysis}

SXL's analysis uses a potential function similar to the one used by~\citet{AroraBlPl98} (henceforth called \defn{ABP}) to bound
the number of steal attempts.  The potential function there is based on the
\defn{enabling tree} --- we say that $u$ enables $v$ if $u$ is the last parent
of $v$ to execute and, in this case, we add an edge between $u$ and $v$ in the
enabling tree.  It turns that, for technical reasons, when using proactive
work stealing, we cannot use the enabling tree.  Instead, we will use the DAG
itself to decide the potential of the node.

The potential function is based on the depth of nodes in the DAG.  The depth
of the node $u$ with one parent $v$ is $d(u) = d(v) + 1$.  The depth of a node
with multiple parents is similar, except that we add 1 to the depth of the
deepest parent.  The weight of node $u$ is $w(u) = \spa - d(u)$.

We say that a node $u$ is the \defn{assigned} node for deque $d$ if $d$ is the
active deque for some worker $p$ and $p$ is currently executing $u$.  The
potential of a node $u$ is defined as follows: $\Phi(u) = 3^{2w(u) - 1}$ if
$u$ is assigned and $\Phi(u) = 3^{2w(u) }$ is ready.  For technical reasons,
we will say that the assigned node for deque $d$ is at the bottom of deque $d$
even though it can not be stolen.  The total potential of a deque $d$ is the
sum of the potential of all nodes on $d$ including the assigned node if $d$ is
active.  The total potential of the computation is the sum of the potentials
of all the ready and assigned nodes on all the deques.

Some of the key results from ABP carry over with these changes in
definitions.
\begin{lemma} 
The initial potential is $3^{2\spa-1}$ and the final potential is
$0$.  In addition, the potential never increases.
\lemlabel{initNoIncr}
\end{lemma}

\begin{lemma} \textbf{Top Heavy Deques}
The top most node in the deque has a constant fraction of the
total potential of the deque.
\lemlabel{topHeavy}
\end{lemma}

The intuition is that the top of the deque contains the node that was
pushed on the deque farthest in the past and therefore, it is the
shallowest node in the DAG.  Since the potential decreases
geometrically with the depth, this node contains most of the potential
of the deque.  

The following lemma is a straightforward generalization of Lemmas 7
and 8 in ABP~\cite{AroraBlPl98}.  The high-level intuition is that
since the top node of each deque contains constant fraction of its
potential, if we steal and execute the top node from each deque with
reasonable probability, the overall potential is likely to reduce by a
constant fraction.
\begin{lemma}
\lemlabel{ABP-gen} Let $\Phi_i$ denote the potential on 
deques at time $t$ and say that the probability of each deque being a
victim of a steal attempt is \emph{at least} $1/X$.  Then after $X$
steal attempts, the potential of deques is at most
$\Phi(t)/4$ with probability at least $1/4$.\qed
\end{lemma}

In ABP, since there are only $P$ deques, one for each worker, this
lemma shows that $P$ random steal attempts reduce the potential by a
constant factor with constant probability.  However, in \ALG, there
are potentially many deques.  Therefore, we may need many more steal
attempts to reduce the potential.  In addition, it is difficult to
design a way to steal from all deques with equal probability if the
deques are distributed across many workers.  

In \ALG, however, recall that that when a deque is suspended, the
worker picks a random worker and donates the deque to that worker.
Therefore, even if one worker suspends many deques, it does not hold
on to them --- the deques are approximately evenly distributed among
all workers.  When a worker steals, it picks a random victim worker
and then a random stealable deque from the victim.  Therefore, each
deque has an approximately equal chance of being a victim of a steal
attempt.  In particular, SXL show the following:

\begin{lemma} Given $P$ workers and $D$ stealable
deques in the system, each worker has at most $D/P + O(\lg P)$
stealable deques with probability at least $1-o(1)$.
\lemlabel{equalDeques}
\end{lemma}

Another insight SXL uses is that a steal attempt from a stealable deque
is generally successful if it is not an active deque.  Therefore, if
there are many (more than $3P$) stealable deques in the system (and
only $P$ of them are active), then most processors have at least one
stealable deque and most steal attempts are successful.  These periods
are called \defn{work-bounded phases} and SXL argue that the total number of
steal attempts in work-bounded phases can be bounded by $O(\work)$ in
expectation.

Therefore, we only need worry about decreasing the potential when
there are not too many stealable deques --- these times are called
\defn{steal-bounded phases}.  SXL use Lemma~\ref{lem:equalDeques} to
argue that each stealable deque during a steal-bounded phase has at
this times has at least $c/P\lg P$ chance of be of being a victim of
a steal attempt (for some constant $c$) since no worker has more than
$O(\lg P)$ stealable deques.  Therefore, using
Lemma~\ref{lem:ABP-gen}, the potential of deques reduces by a
constant factor after $P\lg P$ steal attempts (since unstealable
deques are empty and have no potential).  Given that the initial
potential is $3^{2\spa-1}$, the expected number of steal attempts
during steal bounded phases is $O(P \lg P \spa)$.  Therefore,
considering both work- and steal-bounded phases, the total number of
steal attempts is $O(\work + P \lg P \spa)$.  In addition, they also
separately bound the expected number of successful steals in work
bounded phases is $O(m_k \spa)$.  This allows them to bound the
deviations by $O((P \lg P + m_k) \spa)$.

\paragraph{Changes to potential and analysis to handle weighted edges}

We want to show the same bounds when we use futures for I/O.  The
bounds on steals in work-bounded phases carry over unchanged.  In
particular, the expected number of steal attempts in work-bounded phases
is still $\work$ and the expected number of successful steals is still
$m_k \spa$.  However, for steal bounded phases, where we rely on
potential to bound the number of steal attempts does not apply
directly for somewhat technical reasons.  

Consider the following scenario.  Some worker $p$ with active deque $d$
executes a \getF on an I/O future handle $f$ and blocks since the future is
not ready.  It suspends this deque and steals.  At some point, the \IOT
completes the I/O corresponding to $f$, executes the put, enables the future
join node (say $u$) for $f$, and puts it at the bottom of $d$.  Note that this
deque's potential now suddenly increases, and our analysis strongly depends on
the potential never increasing.  

This is not a problem for SXL for the following reason: If a particular future
join node $u$ is not ready, then some deque must have some ancestor of $u$ on
its deque (either as a ready or an assigned node).  Therefore, $u$ does not
appear on $d$ from nowhere --- some ancestor executes, this ancestor's
potential is larger than $u$'s potential and therefore, even though $u$
becomes ready, the overall potential of the computation does not increase.  In
our case, no ancestor of $u$ is ready or assigned anywhere in the system since
the reason $u$ is not ready is due to the latency on an I/O edge.  This is
problematic since $u$ being enabled increases the potential of the system.  

To fix this problem, we have to give potential to put nodes for I/O futures
(even though they are executed by I/O threads) and handle them in a special
way.  In particular, recall that the only heavy edges in our DAG are between
I/O setup nodes and the corresponding I/O put nodes.  For I/O put nodes, we
will define two notions of depth: the initial depth $id(u)$ of an I/O put node
with enabling parent $v$ (which is always an I/O setup node) is $id(u) =
d(v)+1$.  The depth $d(u)$ starts out as $id(u)$ and reduces on every time
step while the I/O is pending and this I/O put node is suspended.  If the
weight of the heavy edge (the latency of the corresponding I/O) between $v$
and $u$ is $\ell$, then $u$ is suspended for $\ell$ steps.  Therefore, $u$'s
final depth is $fd(u) = d(v) + \ell$.  When the I/O completes, this put node
completes.

Now consider the child node $j$ of the I/O put node --- $j$ is always a future
join node. When deciding the depth of $j$, we always use $fd(u)$.  That is, if
$x$ is $j$'s other parent (the node generated by the \getF operation), then
$d(j) = \max\{fd(u), d(x)\}+1$.  

The potential of a pending put node is defined just like other ready nodes.
At any time, if the depth of the put node is $d(u)$, its weight is $w(u) =
\spa - d(u)$ and potential is $3^{2w(u)}$.  However, since the depth of the
node changes over time, so does its weight and potential.  The total potential
of the computation is the sum of the potentials of all the ready and assigned
nodes on all the deques as well as the potentials of all the put nodes (which
are not on any deque).  

We now get back the following lemma:
\begin{lemma}
The potential never increases.
\lemlabel{noInc}
\end{lemma}
\begin{proof}
We only need consider the case when a future join node $v$ is enabled by a
I/O put node $u$.  By definition, $u$ has lower depth and therefore
higher potential than $v$.  $v$ is only enabled once $u$ finishes.
Therefore, the potential of the system does not increase.  
\end{proof}

However, adding these put nodes creates a problem.  These put nodes are not on
any deque; therefore, steal attempts do not reduce the potential associated
with these nodes directly.  We must also now argue that the potential of I/O
put nodes decreases appropriately during steal-bounded phases.  This is the
reason why we designed the potential of these put nodes in the funny way where
their potential starts out high and reduces on every time step.

\begin{lemma}
During steal-bounded phases, if the total potential at time $i$ is
$\Phi_i$ (including potential of assigned, ready and suspended put
nodes), then after $cP \lg P$ steal attempts, for some constant $c$,
the potential is at most $\Phi(t)/4$ with probability at least $1/4$.
\end{lemma}
\begin{proof}
SXL already argued that the potential of deques reduces appropriately.
Therefore, we only need to consider the suspended I/O put nodes.  Note that if
the latency of a weighted edge is $\ell$, then the corresponding put node
remains suspended for $\ell$ time steps.  Its potential starts at
$3^{2(\spa-id(u))}$ and decreases by a factor of $1/9$ on every time step.
When the I/O completes and the future is ready, the potential of the put node
is $3^{2(\spa-fd(u))}$.  Since it takes at least $c\lg P \geq 1$ time steps
to do $cP \lg P$ steal attempts, the potential of this put node reduces by a
large fraction during this time.  This is true for all put nodes, giving the
desired result.    
\end{proof}

This lemma allows us to bound the expected number of steal attempts
(and therefore, expected number of successful steals) during steal
bounded phases by $P \lg P \spa$ --- the same result as SXL.  Since
the expected number of steal attempts and successful steals for
work-bounded phases remains unchanged, we get the same time and
deviation bounds as SXL.  
\begin{theorem}
The expected number of steal attempts is $O(\work + P \lg P \spa)$.
Therefore, the expected running time is $O(\work + P \lg P \spa)$.
In addition, expected number of deviations is $O((P \lg P + m_k) \spa)$.  
\end{theorem}

\secput{eval}{Empirical Evaluation}

This section empirically evaluates our prototype implementation of \SYS using
a microbenchmark \benchName that closely resembles the example shown in
\secref{system} \figref{example}.  We would like to answer the following three
questions in the evaluation: 1) how well does \SYS hide latency; 2) how much
latency-hiding \SYS can do compared to an ``idealized'' system that hides all
the latency but incurs no additional overhead; and 3) how much each
mechanism used to hide latency contributes to its overhead.  
How we measure each is explained in its respective
subsections.  Overall, the empirical results indicate that, \SYS hides latency
well and can obtain significant speedup compared to $T_1$ running on Cilk-F.
When the latency is short, one could use the oversubscribing strategy
(allocating more workers than number of cores and let the OS scheduling hide
latencies) and obtain benefit.  However, \SYS breaks even compared to the
oversubscription strategy at a latency of $7$ milliseconds using \benchName, and
outperforms the oversubscription strategy after that.  The implementation of
\SYS has been demonstrated to be lightweight, incurring minimal overhead
comparing to an \defn{idealized} version (that incurs zero latency with no
system mechanism overhead).  

\paragraph{Experimental setup.} We ran our experiments on a machine with two
Intel Xeon Gold 6148 processors, each with 20 2.40-GHz cores, with a total of
40 cores.  Each core has a 32 KB L1 data cache, 32 KB L1 instruction cache,
and a 1 MB L2 cache.  Hyperthreading is enabled.  Both sockets have a 27.5 MB
shared L3 cache, and 768 GB of main memory. \SYS and \benchName are compiled
with LLVM/Clang 3.4.1 with \texttt{-O3 -flto} running on Linux kernel version
4.15.  Each data point is the average of $10$ runs.  All data points have
standard deviation less than $5\%$ except for a couple data points at $6\%$.

\paragraph{Benchmark} We use a microbenchmark with very similar code structure
to the map and reduce example (\benchName) described in \secref{system}
\figref{example}), which is also used by \citet{MullerAc16}.
Like Muller and Acar, we emulate $5000$ remote server connections with
simulated delays.  At \liref{connect} in \figref{example}, rather than opening
a true network connection, we used a timed file descriptor which becomes ready
for I/O when the timer expires.\footnote{This functionality is provided by the
Linux \texttt{timerfd}~\citep{LinuxManPageTimerfdCreate19}.}  We replace the
parameter $f$ with a parallel version of the naive recursive implementation of
Fibonacci with a serial base case of 15, and used it to compute the 30th
Fibonacci number.  In place of calling function g (\liref{g}), we return the
sum of \texttt{r1} and \texttt{r2}.

\subsection*{How Well Can \SYS Hide I/O Latencies}

To answer question 1), we compare \SYS with two versions of
Cilk-F:\footnote{Cilk-F extends Cilk Plus to support futures.
\citet{SingerXuLe19} empirically evaluated Cilk-F and showed that it performs
comparably to Cilk Plus.} one uses the same number of workers (Cilk-F) and one
uses twice as many workers so as to oversubscribe the system (Cilk-F (O)) and
let the underlying OS perform scheduling to hide latency.  The \SYS executes
\benchName that uses IO futures to hide latencies whereas the two versions of
Cilk-F execute the baseline code that simply uses blocking \sysRead.  Note
that, since \sysRead is used in place of IO futures, the baseline version
contains only \spawn and \sync.

\begin{figure*}
  \begin{small}
  \begin{tabular}{cc}
    \includegraphics[scale=0.4]{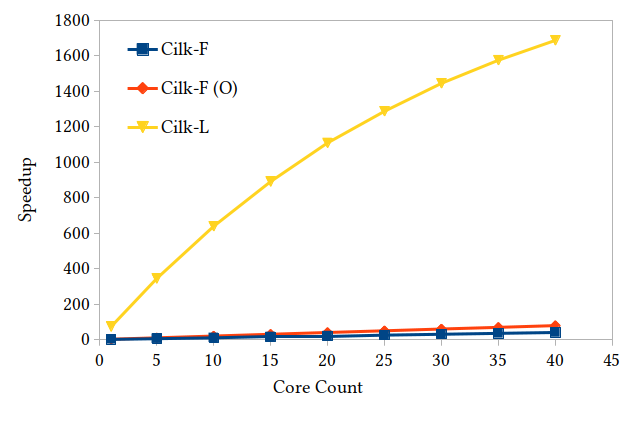} &  
    \includegraphics[scale=0.4]{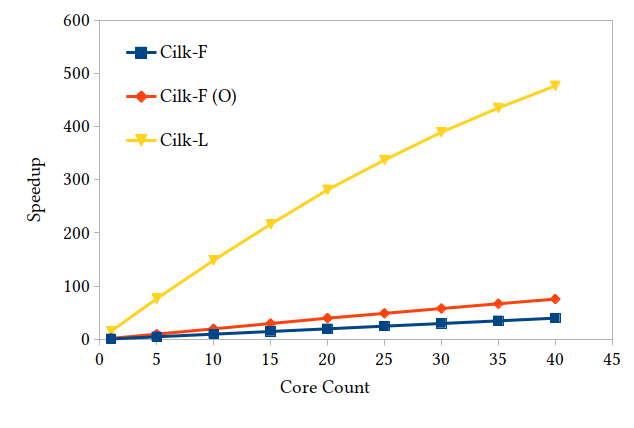}\\ 
    (a) Latency = 500 milliseconds & (b) Latency = 100 milliseconds \\ 
    \includegraphics[scale=0.4]{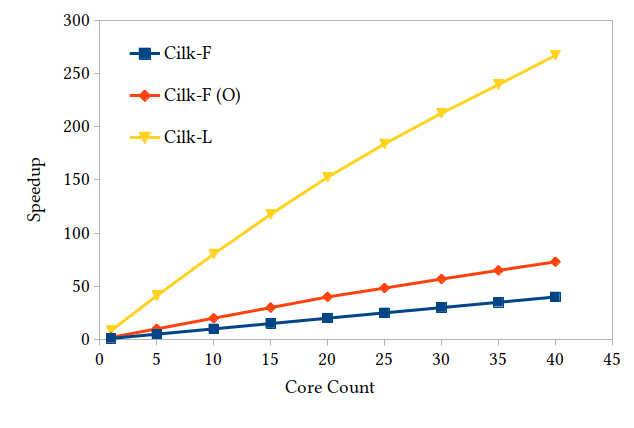} &  
    \includegraphics[scale=0.4]{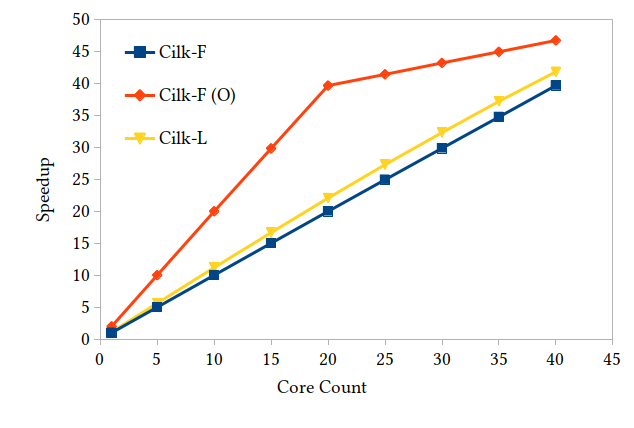} \\ 
    (c) Latency = 50 milliseconds\xspace& 
    (d) Latency = 1 millisecond\xspace\xspace \\ 
  \end{tabular} 
  \caption{Speedups of \benchName running in Cilk-F, Cilk-F (O), and \SYS  
    with latencies of (a) $500$ ms, (b) $100$ ms, (c) $50$ ms, and 
    (d) $1$ ms (where ms is millisecond). 
    The x-axis shows the core counts ($P$) and the y-axis shows the speedup
        compared to the one-core running time on Cilk-F.    
    Cilk-F (O) oversubscribes the system by using $2P$ workers instead of $P$,
        for $P$ number of cores (i.e. every core has two workers pinned, 
        one per hyperthread context).  \SYS pins one worker thread and 
        one \IOT per core, each on its own hyperthread context.}
  \label{fig:perf}
  \end{small}
\end{figure*}

We ran \benchName with simulated I/O latencies of $1$ millisecond, 
   $50$ milliseconds, $100$ milliseconds, and $500$ milliseconds.  
   \figref{perf} shows the speedup of \SYS 
compared to the one-worker execution time of running the baseline version of
\benchName on Cilk-F.  Similar to what was observed by ~\citet{MullerAc16},
we see little advantage to using \SYS to hide I/O latency at $1$ ms.  In fact,
for a latency of $1$ ms, we see that it may be
preferable to utilize both hyperthread contexts as traditional workers rather
than using one for \IOT.  By oversubscribing, at least for \benchName, Cilk-F
can hide latency better than \SYS.  We took separate measurements to see where
the ``break even'' points are, and \SYS and oversubscribing Cilk-F break even
at around latency of $7$ ms.

However, by the time latency hits $50$ ms, there is a significant advantage to
using the I/O functions provided by \SYS. With $P=1$, we already see a speedup
greater than $8\times$, which increases to over $266\times$ at $P=40$.  On the
other hand, oversubscribing Cilk-F achieves less than $2P\times$ speedup.
This pattern continues as we increase the latency to $100$ ms and $500$ ms,
reaching speedups greater than $476\times$ and $1685\times$ respectively. 

One reason why we chose this benchmark and these latency configurations is so
that we could indirectly compare to prior work by \citet{MullerAc16}.
They measured their latency hiding prototype
implementation in parallel ML using similar experimental setup and were only
able to achieve speedups around $80\times$ with the latency of $500$ ms.  Note
that, however, this indirect comparison may is not apple-to-apple, since their
implementation is based on parallel ML, which has its inherent overhead in
memory management. 

\begin{figure*}
  \centering
  \begin{footnotesize}
    \setlength\tabcolsep{2pt}
    \newcommand{\hdr}[1]{\multicolumn{1}{c}{\textit{#1}}}
    \newcommand{\oh}[1]{\hspace{0.5pt}\scriptsize (#1$\times$)}
    \begin{tabular}{c|ccccccccc}
      \textit{latency} & \hdr{$T_1$} & \hdr{$T_5$} & \hdr{$T_{10}$} & 
      \hdr{$T_{15}$} & \hdr{$T_{20}$} & \hdr{$T_{25}$} & \hdr{$T_{30}$} & 
      \hdr{$T_{35}$} & \hdr{$T_{40}$} \\ 
      \hline 
      \textit{ideal} & 33.08 \oh{1.00} & 6.63 \oh{1.00} & 3.31 \oh{1.00} & 2.20 
      \oh{1.00}  & 1.65 \oh{1.00} & 1.33 \oh{1.00} & 1.11 \oh{1.00} & 0.95 
      \oh{1.00} 
      & 0.84 \oh{1.00} \\ 
      \textit{1 ms} & 33.76 \oh{1.02} & 6.77 \oh{1.02} & 3.40 \oh{1.03} & 2.28 
      \oh{1.04} & 1.72 \oh{1.04} & 1.39 \oh{1.05} & 1.18 \oh{1.06} & 1.02 
      \oh{1.07} & 0.91 \oh{1.08} \\ 
      \textit{50 ms} & 33.89 \oh{1.02} & 6.89 \oh{1.04} & 3.53 \oh{1.07} & 2.41 
      \oh{1.10} & 1.86 \oh{1.13} & 1.54 \oh{1.16} & 1.33 \oh{1.20} & 1.18 
      \oh{1.24} & 1.06 \oh{1.26} \\ 
      \textit{100 ms} & 33.99 \oh{1.03} & 6.96 \oh{1.05} & 3.58 \oh{1.08} & 
      2.46 \oh{1.12} & 1.90 \oh{1.15} & 1.58 \oh{1.19} & 1.37 \oh{1.23} & 1.23 
      \oh{1.29} & 1.12 \oh{1.33}\\ 
      \textit{500 ms} & 34.53 \oh{1.04} & 7.36 \oh{1.11} & 3.97 \oh{1.20} & 
      2.85 \oh{1.30} & 2.29 \oh{1.39} & 1.97 \oh{1.48} & 1.76 \oh{1.59} & 1.61 
      \oh{1.69} & 1.50 \oh{1.79} \\
    \end{tabular}
    \caption{The execution times, in seconds, of \benchName using \SYS with 
      different latencies (ms is milliseconds). The values in parentheses 
      are overheads relative to the corresponding $T_P$ time of 
      \textit{ideal}, which runs the baseline of \benchName on Cilk-F 
      with (effectively) zero latency and incurs no system overhead 
      for latency hiding.}
    \label{fig:overhead}
  \end{footnotesize}
\end{figure*}

\subsection*{\SYS's Proximity to Ideal} 

Now we evaluate how close \SYS is to an ``idealized'' version at hiding I/O
latencies.  We obtain the \defn{ideal} measurement by running Cilk-F with a
timed file descriptor with zero delay.\footnote{Technically we used $1$
nanosecond latency, which is the smallest latency one could specify with 
the timed file descriptor on Linux, but it effectively causes the read 
to become ready immediately.}
Moreover, since it is run with Cilk-F, it also does not incur any overhead 
of setting up IO futures, \epoll, nor waking up and context switching to \IOTS. 

\figref{perf} shows the raw execution time of the ideal version and 
that of \SYS with different latencies.  The overhead ($T_P$ of \SYS divided
by $T_P$ of ideal) starts out small with small $P$ and increases as $P$ gets
larger.  
This is in part due to the fact that, the relative ratio between I/O latencies
and compute time increases as the latency increases.  By measuring
\texttt{fib} of $30$ running on one worker, the total amount of work is about
$100$ milliseconds, which is small compared to the $500$ millisecond latency. 
As the theoretical bound predicts, if there is high latency on the span, 
the execution time could be dominated by the span.  We believe this is
what happens in the case of $500$ millisecond latency with high number of
cores.  We confirm this hypothesis by running the same experiments with
\texttt{fib} of $35$, and the gap between \SYS with latency $500$ ms on $40$
cores decreases to be within $10\%$.

\begin{figure*}
  \centering
  \begin{footnotesize}
    \setlength\tabcolsep{2pt}
    \newcommand{\hdr}[1]{\multicolumn{1}{c}{\textit{#1}}}
    \newcommand{\oh}[1]{\hspace{0.5pt}\scriptsize (#1$\times$)}
    \begin{tabular}{c|ccccccccc}
      \textit{overhead} & \hdr{$T_1$} & \hdr{$T_5$} & \hdr{$T_{10}$} &  
      \hdr{$T_{15}$} & \hdr{$T_{20}$} & \hdr{$T_{25}$} & \hdr{$T_{30}$} & 
      \hdr{$T_{35}$} & \hdr{$T_{40}$} \\ 
      \hline 
      \textit{ideal} & 33.08 \oh{1.00} & 6.63 \oh{1.00} & 3.31 \oh{1.00} & 2.20 
      \oh{1.00} & 1.65 \oh{1.00} & 1.33 \oh{1.00} & 1.11 \oh{1.00} & 0.95 
      \oh{1.00} & 
      0.84 \oh{1.00} \\
      \textit{+future} & 32.28 \oh{0.98} & 6.45 \oh{0.97} & 3.22 \oh{0.97} & 
      2.15 \oh{0.98} & 1.61 \oh{0.98} & 1.29 \oh{0.97} & 1.08 \oh{0.97} & 0.93 
      \oh{0.98} & 0.81 \oh{0.96} \\ 
      \textit{+epoll} & 33.70 \oh{1.02} & 6.74 \oh{1.02} & 3.37 \oh{1.02} & 
      2.25 
      \oh{1.02} & 1.69 
      \oh{1.02} & 1.35 \oh{1.02} & 1.13 \oh{1.02} & 0.97 \oh{1.02} & 0.85 
      \oh{1.01}\\ 
      \textit{+IO Thread} & 33.72 \oh{1.02} & 6.76 \oh{1.02} & 3.39 \oh{1.02} & 
      2.28 \oh{1.04} & 1.71 \oh{1.04} & 1.39 \oh{1.05} & 1.18 \oh{1.06} & 1.02 
      \oh{1.07} & 0.91 \oh{1.08} \\ 
    \end{tabular} 
    \caption{The execution times, in seconds, of \benchName with various 
      configurations of \SYS with no I/O latency (i.e. reads do not block). 
      The overheads are relative to the corresponding \textit{$T_P$} time of 
      \textit{ideal}, which uses Cilk-F without latency-hiding. }
    \label{fig:overhead_breakdown}
  \end{footnotesize}
\end{figure*}

\subsection*{Overhead in Latency-Hiding}

The use of IO futures in \SYS has some inherent overhead: 1) setting up and
tear down of IO futures, 2) invoking the \epoll mechanism, which has its
inherent system call overheads, and 3) waking up and context switching into
\IOTS.  Likely these overheads contribute to both the less preferable
performance comparing to oversubscribing Cilk-F when the latency is small and
the additional overhead comparing to the ideal version.  To figure out how
much overhead contributed by each source, we measure different versions of
\SYS and compare that to the ideal version (Cilk-F running the baseline
\benchName with $1$ nanosecond latency).  The \texttt{+future} version is
similar to \texttt{ideal} except with the overhead of using IO futures
(\createF, placing the result into future handles, and \getF).
Building on the \textit{+future} version, the \textit{+epoll} version then
adds the overhead of using \epoll.  Finally, the \textit{+IO Thread} version
adds the overhead of using a separate thread to handle the I/O.  Note that,
however, since the latency is effectively zero, the \IOT will be woken up
only once per request when the request is inserted into \commQ.
\figref{overhead} shows the comparison.  The empirical results show that, the
overhead from use of futures is negligible.  The overhead from 
\epoll and \IOT are comparable, but both are small.

\punt{
Overall, we conclude that, the implementation of \SYS is fairly lightweight
and can hide latency well.  However, if the latency is expected to be very
short (less than $7$ millisecond), then it may be preferable to use standard
blocking I/O (i.e., \sysRead and \sysWrite).
}

\secput{related}{Related Work}

\paragraph{Interesting Use of Futures:}
Since its proposal in the late 70th~\cite{FriedmanWi78, BakerHe77},
the use of futures has been incorporated into various task parallel
platforms~\cite{ChandraGuHe94, KranzHaMo89, Halstead85, CharlesGrSa05,
  SpoonhowerBlHa08, FluetRaRe10, TasirlarSa11, CaveZhSh11, LuJiSc14}.
Futures are typically used as a high-level synchronization construct
to allow parallel tasks to coordinate with one another in a way that
is more flexible than pure fork-join parallelism.

Researchers have proposed interesting uses of futures.  ~\citet{BlellochRe97} used futures to generate
``non-linear'' pipeline.  Using futures to pipeline the split and
merge of binary trees, they developed a parallel algorithm of tree
merge with better span than a fork-join parallel marge
algorithm. ~\citet{SurendranSa16b} proposed using
futures to automatically parallelize pure function calls in programs
and developed the corresponding compiler analyses. ~\citet{KoganHe14} described a use of futures in the concurrent
programming setting, called \defn{linearizable futures}, that allows a
concurrent data structure to be shared among threads via the use of
futures and formalized the correctness guarantees for such use.
~\citet{MilmanKoLe18} proposed an algorithm for
batched lock-free queue concurrent data structure using futures.

\paragraph{Supporting synchronization primitives in work-stealing:} 
Researchers have also proposed runtime schedulers for scheduling
programs with blocking synchronization.  For instance,~\citet{AgrawalLeSu10a} proposed a runtime system for helper
locks where when a worker tries to get a lock which is not available,
it tries to help complete the critical section that is currently
holding the lock.  They proved that this scheduler was efficient if
large critical sections had sufficient internal
parallelism.  

X10~\cite{CharlesGrSa05} and Habanero~\cite{CaveZhSh11} variants
support synchronization primitives such as conditional blocks, clocks
and phasers are supported.  Most of these implementations do not have
provably efficient performance bounds.
Initially, in X10~\cite{CharlesGrSa05} and Habanero Java,
synchronization primitives (e.g., conditional atomic blocks or
barriers) may cause the worker to simply block, and the runtime
compensated by creating a new worker thread to replace the blocked
worker.  Later, ~\citet{TardieuWaLi12} proposed better
compiler and runtime support for X10 for suspending a task blocked on
synchronizations.  However, the suspended tasks are stored in a
centralized queue.  For Habanero Java,~\cite{ImamSa14a} describe an alternative support: when
suspended tasks become resumable, they are pushed onto the deque of
the worker that executed the operation to unblock the tasks.  

~\citet{ZakianZaKu15} extend Intel Cilk
Plus~\cite{IntelCilkPlus13} to provide support for a low-level library
which allows a worker to suspend the current execution context upon
encountering a blocking I/O and find something else to do.  In this
case, the multiple suspended contexts (deques) are stored with the
worker which suspended them.  However, only the active deque is
exposed to be stolen from.  Thus, a worker may end up with many
suspended deques with high potential nodes that cannot be stolen.

\paragraph{Work-stealing schedulers with multiple deques per worker:}
Various work-stealing runtime systems have used multiple deques per
worker for different reasons.  The runtime system for helper
locks~\cite{AgrawalLeSu10a} (discussed above) used multiple deques per
worker.  When a worker is blocked on a lock, it is only allowed to
work on the critical section, that is holding the lock (assuming this
critical section has internal parallelism) and does so by allocating
another deque specifically for this critical section $C$.  Therefore,
in a program with nested locks with nesting depth $d$, workers could
have as many as $d$ deques each.  However, the scheduler is designed
so that each worker can steal from at most one deque of each of the
other workers.  In a similar vein,~\citet{AgrawalFiLu14}
proposed Batcher runtime system to handle parallel programs that make
data structure accesses.  In this case, workers can be working on
either the program work or the data structure work and this work is
kept on different deques.  But again, at any given time, a worker
steals randomly among $P$ deques.

The closest work to this work is Porridge processor-oblivious record
and replay system for dynamic multithreaded programs using work
stealing~\cite{UtterbackAgLe17}.  Porridge allows multithreaded
program with locks to be executed on some number of processors while
recording all the happens-before relationship between critical
sections.  Later the execution can be replayed on a different number
of processors but guarantees the same happens-before relationships.
During record, the vanilla workstealing algorithm that just blocks on
un unavailable lock can be used.  However, during replay, a vanilla
work-stealing scheduler can lead to deadlocks.  Therefore, if a
critical section $c$ tries to acquire a lock and can not acquire it
since the critical section with a happens-before edge to $c$ has not
finished, the processor must find something else to do.  The runtime
system there also uses proactive work-stealing and achieves similar
bounds.  

\secput{conclusion}{Conclusion}

In order to support modern desktop and server software, I/O operations
should be supported as a fundamental component of task parallel
platforms.  In this paper, we show how one one may incorporate I/O
into a task parallel platform seamlessly with efficient scheduling to
hide I/O latencies.  In particular, our platform, \SYS, provides a
programming API for performing I/O that works harmoneously with
existing parallel control constructs.  In addition, the underlying
runtime system efficiently schedules both the computation and the I/O
operations to provide nearly optimal execution time guarantee and a
bound on the number of deviations.  We achieve this by using proactive
work stealing schedulers recently developed for scheduling
computations with futures.  Empirical evaluation of our prototype
system shows that, I/O can be supported efficiently with effective
latency hiding.

\clearpage
\bibliographystyle{ACM-Reference-Format}
\bibliography{papers}

%

\end{document}